\documentstyle[preprint,aps,prd]{revtex}
\begin{document}
\title{Quantum fluctuations of lightcone  \\in  4-dimensional spacetime with parallel
plane boundaries}
\author{ Hongwei Yu and Pu-Xun Wu }

\address
{ Department of Physics and Institute of  Physics,\\ Hunan Normal
University, Changsha, Hunan 410081, China.}
\tightenlines
\maketitle
\begin{abstract}
Quantum fluctuations of lightcone are examined in a 4-dimensional
spacetime with two parallel planes. Both the Dirichlet and the
Neumann boundary conditions are considered. In all the cases we
have studied, quantum lightcone fluctuations are greater where the
Neumann boundary conditions are imposed, suggesting that quantum
lightcone fluctuations depend not only on the geometry and
topology of the spacetime as has been argued elsewhere but also on
boundary conditions. Our results also show that quantum lightcone
fluctuations are larger here than that in the case of a single
plane. Therefore, the confinement of gravitons in a smaller region
by the presence of a second plane reinforces the quantum
fluctuations and this can be understood as a consequence of the
uncertainty principle.
 \vspace{0.2cm}
\end{abstract}

\newpage

\section{Introduction}


Quantization of gravity is more difficult than quantization of the
other fundamental interactions. Although a lot of efforts have
been made, a consistent theory of quantum gravity is still
elusive.  If, however, the basic quantum principles we are already
familiar with apply as well to a quantum theory of gravity, we can
make some predictions about expected quantum effects, even in the
absence of a fundamental underlying theory. One such effect is the
fluctuations of lightcone induced by quantum fluctuations of the
space-time metric to be expected in any theory of quantum gravity.
 The study of quantum lightcone fluctuations has attracted much attention recently. A model of lightcone
fluctuations on a flat  background has been developed\cite{Ford1},
where the fluctuations are produced by gravitons propagating on
the background. This model has been further
developed\cite{Ford2,yu1} and applied to study the lightcone
fluctuations in spacetimes with a compactified spatial section and
with a single plane boundary\cite{yu1} as well as in spacetimes
with extra dimensions\cite{yu2,yu3}. It has also been applied to a
microscopic recoil model for lightcone fluctuations in a quantum
gravity framework inspired by string theory\cite{EMN99}.
 Because the lightcone fluctuates,
the propagation time of a classical light pulse over distance $r$
is no longer precisely $r$, but undergoes fluctuations around a
mean value of $r$. The quantum lightcone fluctuations can be
thought of  as giving rise to stochastic fluctuations in the speed
of light, which may produce an observable time delay or advance in
the arrival times of pulses from distant astrophysical sources, or
the broadening of spectral lines. They also lead to an intrinsic
uncertainty in measuring positions of objects and thus constitute
a noise from quantum gravity in modern gravity-wave
interferometers. The possible observable distortion of the cosmic
microwave background (CMB) radiation spectrum due to the lightcone
fluctuations has been recently discussed\cite{DFYF00,EMN02}. These
quantum gravity effects as well as others predicted by various
other approaches\cite{GJND,GJNDS,GP,AMU,EFMMN} open up the
possibilty of testing theories of quantum gravity in the near
future or even at the present experiments. This might suggest that
an era of quantum-gravity phenomenlogy may be just around the
corner\cite{GAC}.

In this paper, our aim is to extend the analysis performed in
Ref.\cite{yu1} on lightcone fluctuations in a spacetime with a
single plane boundary to the case where there are two parallel
plane boundaries, and examine what influence of the presence of
the second plane will have on the lightcone fluctuations. Both the
Dirichlet and the Neumann boundary conditions will be considered.
Note that in Ref.\cite{yu1} only the Neumann boundary conditions
were studied for the single plane.  The framework we are going to
use is that developed in \cite{Ford1,yu1,yu2}. In Sect. II, we
will give a brief
 review of the formalism and   then study in detail the lightcone
 fluctuations for both
longitudinal and transverse light propagations in both cases of
the Dirichlet and the Neumann boundary conditions. Our results
will be summarized in Sect. III.


\section{Lightcone fluctuations in spacetime with two parallel plane boundaries}


 To begin, let us
 consider a flat background  spacetime  with a linearized perturbation
$h_{\mu\nu}$ propagating upon it, so the spacetime metric may be
written as
 \begin{equation}
ds^2  = (\eta_{\mu\nu} +h_{\mu\nu})dx^\mu dx^\nu
= dt^2 -d{\bf x}^2 + h_{\mu\nu}dx^\mu dx^\nu \, ,
 \end{equation}
where the indices $\mu,\nu$ run through $0,1,2,3$.
Let $\sigma(x,x')$ be one half of the squared geodesic distance between
 a pair of spacetime points $x$ and $x'$,   and $\sigma_0(x,x')$
be the corresponding quantity in the flat background.
In the presence of a linearized metric  perturbation
 $h_{\mu\nu}$, we may expand
 \begin{equation}
\sigma = \sigma_0 + \sigma_1 + O(h^2_{\mu\nu}) \; .
\end{equation} Here $\sigma_1$ is first order  in $h_{\mu\nu}$.
If we quantize $h_{\mu\nu}$, then quantum gravitational vacuum
fluctuations will lead to fluctuations in the geodesic separation,
and therefore induce
 lightcone fluctuations.  In particular, we have
$\langle \sigma_1^2 \rangle \not= 0$, since $\sigma_1$  becomes a quantum
operator when the metric perturbations are quantized.   The quantum lightcone
fluctuations give rise to
stochastic fluctuations in the speed of light, which may produce an observable
time delay or advance $\Delta t$ in the arrival times of pulses. Note that
this model uses a linearized approach to quantum gravity which is expected
to be a limit of a more exact theory. In the absence of a full theory,
this seems to be the most conservative way to compute quantum gravity effects.

Let us  consider the propagation of light pulses between a source
and a detector separated by a distance $r$ on a flat  background
with quantized linear perturbations.  It has been shown that the
root-mean-squared fluctuation in the propagation time is given by
\cite{yu2}
\begin{eqnarray}
\label{eq:1}
\Delta t=\frac{\sqrt{|\langle\sigma_1^2\rangle_R|}} {r}\;,
\end{eqnarray}
where $\langle \sigma_1^2 \rangle_R$ is a renormalized expectation
value. In order to find $\Delta t$ we need to calculate the
quantum expectation value of the mean squared fluctuation
$\langle\sigma_1^2\rangle_R$ of the geodesic interval function,
which is given by\cite{Ford1,yu1}
\begin{eqnarray}
\label{eq:3}
\langle\sigma_1^2\rangle_R=
\frac{1}{8}(\Delta r)^2\int_{r_0}^{r_1}dr
\int_{r_0}^{r_1}dr'n^\mu n^\nu n^\rho n^\sigma G^R_{\mu\nu\rho\sigma}(x,x')\;,
\end{eqnarray}
where $dr=|d{\bf x}|$, $\Delta r=r_1-r_0$, $n^\mu=dx^\mu/dr$, and
$G^R_{\mu\nu\rho\sigma}(x,x')$ is the  suitably renormalized
graviton Hadamard function.  A natural starting point for quantum
calculations may be the graviton two point function of Minkowski
spacetime. If, however, it is not renormalized, the integral
diverges. The usual response to this problem is to require that
the renormalized two point function vanish in Minkowski spacetime.
If this is the correct solution, then $\Delta t = 0$, and there
are no lightcone fluctuations in Minkowski spacetime. It should be
pointed out that many current theories of quantum gravity do not,
however, use 4-dimensional Minkowski spacetime as the starting
point. For example, quantum gravity derived in the context of
string theory \cite{EMN99} and canonical quantum gravity in the
loop approximation \cite{GP} all predict lightcone fluctuations.

From now on, we will assume that a 4-dimensional spacetime
configuration in which there are two two-dimensional
 parallel planes located at $z=0$ and $z=L$ respectively. We will examine the quantum lightcone fluctuations
 in this scenario for both
 the Dirichlet and the Neumann boundary conditions.


 \subsection{Dirichlet boundary conditions}


 First we
 will consider the case where gravitons satisfy the Dirichlet boundary conditions.
Assuming that $G_{ijkl}^{(1)}(t,x,y,z;t',x',y',z')$ is the Hadamard function
for the Minkowski vacuum state\footnote{By Minkowski, we mean flat spacetime without any boundaries}, the
renormalized Hadamard function  for the case of two parallel planes can be found, by using the method of  images, as
\begin{eqnarray}
\label{eq:2}
G^R_{ijkl}(t,x,y,z;t',x',y',z')&=&\sum_{n=-\infty}^{+\infty}{'}G_{ijkl}^{(1)}(t,x,y,z;t',x',y',z'+2nL)\nonumber\\
&&-\sum_{n=-\infty}^{+\infty}G_{ijkl}^{(1)}(t,x,y,z;t',x',y',-z'+2nL)\;.
\end{eqnarray}
Here the prime denotes that the $n=0$ term which is the Hadamard
function of Minkowski spacetime is omitted in the summation. Since
the formalism is gauge invariant\cite{yu1}, we will choose to work
in the TT gauge.


\subsubsection{ Parallel propagation case}


Now let us consider  a light ray traveling parallel to the planes,
along the $x$-direction from point $(t,a,0,z)$ to point
$(t',b,0,z)$, Define $\rho=x-x'$, and $\Delta z=z-z'$ and keep in
mind that the integration in Eq.(\ref{eq:3}) is to be carried out
along the null ray on which $\rho=t-t'$, we obtain the relevant
graviton two-point function in the Minkowski vacuum state in the
TT gauge\cite{yu1}
\begin{eqnarray}
G_{xxxx}^{(1)}(t,x,0,z;t',x',0,z')&=&\frac{2}{\pi^2}\bigg[-\frac{\rho^2\Delta
z^4}{8(\rho^2+\Delta z^2)^4} -\frac{\rho^6}{3(\rho^2+\Delta
z^2)^4}+
\frac{47\rho^4\Delta z^2}{12(\rho^2+\Delta^2)^4}\nonumber\\
&&-\frac{1}{(\rho^2+\Delta z^2)^{9/2}}(-\frac{1}{32}\rho\Delta z^6-\frac{3}{8}\rho^3\Delta z^4\nonumber\\
&&+
\frac{3}{4}\rho^5\Delta z^2)\cdot \ln\left(\frac{\sqrt{\rho^2+\Delta z^2}+
\rho}{\sqrt{\rho^2+\Delta z^2}-\rho}\right)^2\bigg]\;,
\end{eqnarray}
 Let $r=b-a$,  we have
\begin{eqnarray}
\label{eq:6} && \int_a^b dx\int_a^b
dx'G_{xxxx}^{(1)}(t,x,0,z;t',x',0,z')=2\int_0^r d\rho
(r-\rho)G_{xxxx}(\rho, \Delta z) \nonumber\\
&&\quad\quad=\frac{2}{\pi^2}\left[-\frac{2r^4+\Delta
z^2r^2}{4(r^2+\Delta z^2)^2}+\frac{8r^5+8r^3\Delta z^2+
3r\Delta z^4}{24(r^2+\Delta z^2)^{5/2}}\ln\frac{\sqrt{r^2+\Delta z^2}+r}{\sqrt{r^2+\Delta z^2}-r}\right]\nonumber\\
&&\quad\quad\equiv f(\Delta z, r)\;.
\end{eqnarray}
Plugging the above result into  Eq.~(\ref{eq:2}) and Eq.~(
\ref{eq:3}) and letting $z=z'$ yields
\begin{eqnarray}
\label{eq:7}
\langle\sigma_1^2\rangle_R &=& \frac{r^2}{8}\left[-f(2z,r)+\sum_{n=1}^{+\infty}(2f(2nL,r)-f(2z-2nL,r)
-f(2z+2nL,r))\right]\nonumber\\
&=&\frac{r^2}{8}\left[-f(2z,r)-f(2z-2L,r)+g\right]
\;,
\end{eqnarray}
where we have defined
\begin{eqnarray}
g=\sum_{n=1}^{+\infty}[2f(2nL,r)
-f(2z+2nL,r)]-\sum_{n=2}^{+\infty}f(2z-2nL,r)\;.
\end{eqnarray}
Since the first term in Eq.(\ref{eq:7}) diverges when
$z\rightarrow 0$, and  the second term blows up  as $z$ approaches
$L$,  the mean squared geodesic interval fluctuation is
ill-behaved on the boundaries. This might be a result of our
assumption that the boundaries are rigid and at fixed positions. A
similar example is the divergence of energy density of a quantized
field as the boundary is approached, where it has been shown if
one treats the boundaries as quantum objects with a nonzero
position uncertainty, the singularity in energy density is
removed\cite{Ford4}. However it remains to be checked whether the
divergence here can also be eliminated when the quantum
fluctuations of the boundaries are taken into account.

Finding a closed-form result for the summations in function $g$
seems to be a very difficult task.  We will not make the attempt
in this paper, instead we will try to analyze two special cases,
i.e., $\epsilon=r/L\gg 1$ and $\epsilon=r/L\ll 1$.  When the
distance travelled by the light ray is much larger than the
separation of the two parallel planes such that $\epsilon\gg 1$,
or $r\gg L$,  the summation in Eq.~(\ref{eq:7}) can be
approximated by integration, for example,
\begin{eqnarray}
\sum_{n=1}^{+\infty}f(2nL,r)&\approx&
\frac{\epsilon}{\pi^2}\int_{1/\epsilon}^\infty dx\left[
-\frac{2+4x^2}{2(1+4x^2)^2}+\frac{8+32x^2+48x^4}{12(1+4x^2)^{5/2}}
\ln\frac{\sqrt{1+4x^2}+1}{\sqrt{1+4x^2}-1}\right]\nonumber\\
&\equiv& \frac{\epsilon}{\pi^2}\;h(1/\epsilon)\;.
\end{eqnarray}
It then follows that
\begin{eqnarray}
g&\approx & \frac{\epsilon}{\pi^2}\left[ 2h(1/\epsilon)-h(1/\epsilon+z/r)-h(2/\epsilon-z/r)\right]  \;.
\end{eqnarray}
Performing the integration by parts and series expanding the
result to order $o(1/\epsilon)$, we obtain
\begin{eqnarray}
g\approx \frac{4}{3\pi^2}\left[\ln\epsilon
-\left(1+\frac{z}{L}\right)\ln\left(1+{z\over L}\right)-
\left(2-\frac{z}{L}\right)\ln\left(2-{z\over L}\right)\right] \;.
\end{eqnarray}
Since $z\leq L$,  we must have $\lambda\gg 1$ when $\epsilon\gg
1$.  Therefore, similarly, one has
\begin{eqnarray}
\label{eq:f}
f(2z,r)\approx \frac{4}{3\pi^2}\ln\frac{r}{z}\;,
\end{eqnarray}
\begin{eqnarray}
\label{eq:f1}
f(2z-2L,r)\approx \frac{4}{3\pi^2}\ln\frac{r}{L-z}\;.
\end{eqnarray}
Hence,  in the case of $r\gg L$
\begin{eqnarray}
\langle\sigma_1^2\rangle_R &\approx
&-\frac{r^2}{6\pi^2}\ln\frac{rL}{z(L-z)} \;,
\end{eqnarray}
and the mean deviation from the classical propagation time is
\begin{eqnarray}
\label{eq:t}
\Delta t\approx \sqrt{\frac{1}{6\pi^2}\ln\frac{rL}{z(L-z)}}\;.
\end{eqnarray}

A few comments are now in order. This result reveals that the mean
deviation due to lightcone fluctuations grows as $r$, the travel
distance, increases, and it is symmetric under $z\leftrightarrow
L-z$ as it should be. At the same time, one can see that the
deviation increases as $L$ decreases. This indicates that the
presence of a second plane reinforces the lightcone fluctuations.
In fact, if we expand Eq.(\ref{eq:t}) in the limit of large $L$,
the result is
\begin{equation}
\label{eq:LL1}
\Delta t\approx \sqrt{\frac{1}{6\pi^2}\ln\frac{r}{z}}+{z\over L}+{1\over 2} {z^2\over L^2}\;,
\end{equation}
 which gives us a idea what the contributions are due to the presence of the second brane in this case. Finally, let us note that the mean
 deviation has a minimum at $z=L/2$.

If $\epsilon \ll 1$, i.e.  $L\gg r$,  we can series expand $g$ and $f(2z-2L,r)$ with $r/L$ to get
\begin{eqnarray}
g+f(2z-2L,r)\approx \frac{2}{\pi^2}\sum_{n=1}^{\infty}\left[\frac{r^6}{90n^6L^6}-\frac{r^6}{180(nL+z)^6}
-\frac{r^6}{180(nL-z)^6}\right]\;.
\end{eqnarray}
The summation can be calculated  and the result can be expanded as
a power series of  $z/L$ to yield
\begin{eqnarray}
g+f(2z-2L,r)\approx
-\frac{\pi^6}{10\times 45^2}\;\left(\frac{r}{L}\right)^6\;\left(\frac{z}{L}\right)^2\;.
\end{eqnarray}
Meanwhile  $f(2z,r)$ is given by Eq.(\ref{eq:f}) if $r\gg z$, and  $f(2z,r)\approx 0$ if $r\ll z$.
Consequently, one has for the mean deviation in the case of $L\gg r$ and $r\gg z$
\begin{equation}
\langle\sigma_1^2\rangle_R\approx-\frac{r^2\ln(r/z)}{6\pi^2}-\frac{r^2\;\pi^6}{80\times
45^2}\;\left(\frac{r}{L}\right)^6\;\left(\frac{z}{L}\right)^2\;
\end{equation}
and
\begin{eqnarray}
\label{eq:tttt} \Delta
t=\frac{\sqrt{|\langle\sigma_1^2\rangle_R|}} {r}\approx
\sqrt{\frac{\ln(r/z)}{6\pi^2}+\frac{\pi^6}{80\times
45^2}\;\left(\frac{r}{L}\right)^6\;\left(\frac{z}{L}\right)^2}\;.
\end{eqnarray}
  And in the case of $L\gg r$ and $r\ll z$, the mean deviation is given instead by
 \begin{eqnarray}
\label{eq:t5} \Delta t\approx {\pi^3\over
180\sqrt{5}}\;\left(\frac{r}{L}\right)^3\;\left(\frac{z}{L}\right)\;.
\end{eqnarray}
These results also show enhancing  lightcone fluctuations as the
travel distance increases or the separation of the plane
decreases.

\subsubsection{Perpendicular propagation case}


Let us turn our attention to the case where  the light ray
propagates perpendicularly to  the plane, from point $(t,0,0,a)$
to $(t',0,0,b)$. Rewrite  Eq.~(\ref{eq:2})  as
\begin{eqnarray}
\label{eq:gz}
G^{R}_{zzzz}(t,z;t',z')&=&-G^{(1)}_{zzzz}(t,z;t',-z')+\sum_{n=-\infty}^{+\infty}{'}[G^{(1)}_{zzzz}(t,z;t',z'+2nL)\nonumber\\
&&-G^{(1)}_{zzzz}(t,z;t',-z'+2nL)]\;.
\end{eqnarray}
Here the prime on the summation indicates that the $n=0$ term is
excluded and the the notation $(t,0,0,z)\equiv (t,z)$ has been
adopted.  Then we  have \cite{yu1}
\begin{eqnarray}
\label{eq:z1}
G^{(1)}_{zzzz}(t,z;t',z')&=&-\frac{2}{\pi^2}\bigg[\frac{\Delta
t^2}{\Delta z^4}+
\frac{\Delta t^3}{4\Delta z^5}\ln\left(\frac{\Delta z-\Delta t}{\Delta z+\Delta t}\right)^2\nonumber\\
&&-\frac{2}{3\Delta z^2}-\frac{\Delta t}{4\Delta
z^3}\ln\left(\frac{\Delta z -\Delta t}{\Delta z+\Delta
t}\right)^2\bigg]\;,
\end{eqnarray}
where $\Delta t=t-t'$ and $\Delta z=z-z'$. Taking into account
that for a null geodesic $\Delta t=\Delta z$ and as before letting
$r=b-a$,  we find, after performing the integrations
\begin{eqnarray}
\label{eq:int1} \int_a^b & dz&\int_a^b
dz'G^{(1)}_{zzzz}(t,z;t',-z')\nonumber\\
&&=\frac{r}{6\pi^2(2a + r)^3}\left\{(4ar+2r^2)-
 (6a^2+6ar+r^2)\ln\frac{a^2}{(a + r)^2}\right\}\;,
\end{eqnarray}
\begin{eqnarray}
\label{eq:int} &&\sum_{n=-\infty}^{+\infty}{'}\;\int_a^b
dz\int_a^b dz'\;G^{(1)}_{zzzz}(t,z;t',z'+2nL)
\nonumber\\&&\quad\quad=\frac{2}{3\pi^2}
 \sum_{n=1}^{+\infty}\bigg[\frac{(r+nL)^3}{(r+2nL)^3}\ln\left(1+\frac{r} {nL}\right)^2+\frac{(r-nL)^3}
 {(r-2nL)^3}\ln\left(1-\frac{r}{nL}\right)^2\nonumber\\
 && \quad\quad+\frac{4r^2(r^2-2n^2L^2)}{(r^2-4n^2L^2)^2}\bigg]\;,
\end{eqnarray}
and
\begin{eqnarray}
\label{eq:int2}
 &&\sum_{n=-\infty}^{+\infty}{'}\;\int_a^b
dz\int_a^b dz'\;G^{(1)}_{zzzz}(t,z;t',-z'+2nL)
\nonumber\\&&\quad\quad=\frac{1}{3\pi^2}
 \sum_{n=1}^{+\infty}\bigg[
 \frac{r[6a^2+6n^2L^2-6nLr+r^2+6a(r-2nL)]}
 {2(2a-2nL+r)^3}\ln\left(\frac{a-nL}
{a-nL+r}\right)^2\nonumber\\
&&\quad\quad+\frac{r^2}{(2a-2nL+r)^2}\bigg]\;,
\end{eqnarray}
Since here $r$ is always less than $L$, we will only consider the
case where $L\gg r$. Again Eq.(\ref{eq:int}) and
Eq.(\ref{eq:int2}) can be approximately evaluated, by expanding
the summands as a power series of $r/L$, keeping only the leading
term and doing the summation afterwards,
\begin{eqnarray}
\label{eq:26}
 \sum_{n=-\infty}^{+\infty}{'}\;\int_a^b dz\int_a^b
dz'\;G^{(1)}_{zzzz}(t,z;t',z'+2nL)
\approx\sum_{n=1}^{\infty}{2r^2\over 3\pi^2n^2L^2}=
\frac{r^2}{9L^2}\;,
\end{eqnarray}

\begin{eqnarray}
\label{eq:27}
 &&\sum_{n=-\infty}^{+\infty}{'}\;\int_a^b dz\int_a^b
dz'\;G^{(1)}_{zzzz}(t,z;t',-z'+2nL)\nonumber\\
\quad \quad &&\approx\sum_{n=-\infty}^{+\infty}{'}{r^2\over
3\pi^2(a-nL)^2} ={1\over 3}{r^2\over L^2}\csc^2(\pi a/L)-{r^2\over
3\pi^2a^2}\;.
\end{eqnarray}
Here we have used the following two formulas in calculating the
summation\cite{Jo}
 \begin{equation} \sum_{n=0}^{\infty}\;{1\over
(n^2-a^2)^2}={1\over 2a^4}+{\pi \over 4a^3}\cot(\pi a)+{\pi^2\over
4a^2}\csc^2(\pi a)\;,
\end{equation}
and
\begin{equation}
\sum_{n=0}^{\infty}\;{n^2\over (n^2-a^2)^2}=-{\pi \over
4a}\cot(\pi a)+{\pi^2\over 4}\csc^2(\pi a)\;.
\end{equation}
 As for  Eq.~(\ref{eq:int1}), we need to consider two special cases. One is when $r\ll a$,  where one finds
\begin{eqnarray}
\label{eq:30}
\int_a^b dz\int_a^b
dz'G^{(1)}_{zzzz}(t,z;t',-z')\simeq \frac{r^2}{3\pi^2a^2}\;,
\end{eqnarray}
and the other is when $r\gg a$ and there one has
\begin{eqnarray}
\label{eq:31}
 \int_a^b dz\int_a^b
dz'G^{(1)}_{zzzz}(t,z;t',-z')\simeq \frac{1}{3\pi^2}[1+\ln (r/a)]
\;.
\end{eqnarray}
To summarize, when $L\gg r$ and $r\gg a$, we have that
\begin{eqnarray}
\label{eq:int3}
\langle\sigma_1^2\rangle_R&=&{r^2\over 8}\;\int_a^b dz\int_a^b dz'G^{R}_{zzzz}(t,z;t',z')\nonumber\\
&&\quad\quad\approx -\frac{1}{24\pi^2}[1+\ln
(r/a)]+\frac{r^2}{72L^2} +{r^2\over 24\pi^2a^2}-{1\over
24}{r^2\over
L^2}\csc^2(\pi a/L)\nonumber\\
&&\quad\quad\approx -\frac{1}{24\pi^2}[1+\ln (r/a)]-{\pi^2\over
360}\left({r\over L}\right)^2 \left({a\over L}\right)^2\;.
\end{eqnarray}
Therefore, the mean deviation due to lightcone fluctuations is
\begin{eqnarray}
\label{eq:T1} \Delta t=\frac{\sqrt{|\langle\sigma_1^2\rangle_R|}}
{r}\approx\sqrt{\frac{1}{24\pi^2}[1+\ln (r/a)]+{\pi^2\over
360}\left({r\over L}\right)^2 \left({a\over L}\right)^2}\;.
\end{eqnarray}
The last term in square-root represents the contribution due to
the presence of a second plane. However, when $L\gg r$ and $r\ll
a$, Eq.~(\ref{eq:int3}) becomes
\begin{eqnarray}
\langle\sigma_1^2\rangle_R&=&{r^2\over 8}\;\int_a^b dz\int_a^b dz'G^{R}_{zzzz}(t,z;t',z')\nonumber\\
&&\quad\quad\approx -{1\over 24}{r^2\over L^2}\csc^2(\pi
a/L)+{1\over 72}\frac{r^2}{L^2} \;.
\end{eqnarray}
Consequently, the mean deviation reads
\begin{eqnarray}
\Delta t=\frac{\sqrt{|\langle\sigma_1^2\rangle_R|}}
{r}\approx\sqrt{ {1\over 24}{r^2\over L^2}\csc^2(\pi a/L)-{1\over
72}\frac{r^2}{L^2} }.
\end{eqnarray}
If we further assume that $ L\gg a$, then the above result becomes
\begin{eqnarray}
\Delta t\approx \sqrt{\frac{r^2}{24\pi^2a^2} +{\pi^2\over
360}\left({r\over L}\right)^2 \left({a\over L}\right)^2}\;.
\end{eqnarray}
This reveals  a linear growth over the travel distance, $r$.

A few comments on the results obtained above are in order here.
Let us recall that $\langle\sigma_1^2\rangle_R$ was assumed to be
positive when the relation between $\Delta t$ and
$\langle\sigma_1^2\rangle_R$ was first derived \cite{Ford1,yu1}.
Thus, if $\langle\sigma_1^2\rangle_R$ is negative, then  $\Delta
t$ would be imaginary.  To solve this problem, an alternative
derivation was proposed based upon averaging $\sigma^4$ over a
given quantum state of gravitons and applying Wick's theorem
\cite{yu2}. This derivation yielded Eq.~(\ref{eq:1}) in which the
absolute value of $\langle\sigma_1^2\rangle_R$ is used. All the
calculations performed so far on the mean deviation in propagation
time, $\Delta t$, are based on this new formula.  Although
Eq.~(\ref{eq:1}) can formally deal with situations where
$\langle\sigma_1^2\rangle_R$ is negative and gives a real thus
physically meaningful $\Delta t$, it does not fully resolve the
issue physically.  It just circumvents the sign problem and tells
us little about the physical significance of the sign in the
geodesic distance fluctuation.  For all the cases we have
discussed above in the case of the Dirichlet boundary conditions,
the mean squared fluctuations in the geodesic distance are all
negative, while, as we will see in the next subsection, they are
all positive in the corresponding cases in the case of the Neumann
boundary conditions.  Now let us note that to build a reflecting
wall for gravitons one would need negative mass. Therefore
imposing the Dirichlet boundary conditions for gravitons in the
present context seems unphysical, as it has the same effect as
having sources of negative mass. This might suggest that the
negative sign for the geodesic distance fluctuation might be a
result of the imposition of the unphysical boundary conditions.
However, it is worth pointing out that much more work needs to be
done to fully understand the meaning of the sign in the geodesic
distance fluctuation and its relation with boundary conditions,
geometry and topology of spacetime as well as other physical
conditions. Finally, recall that our renormalization scheme is to
subtract the mean squared fluctuation of the geodesic interval
function in the Minkowski spacetime. Thus our result may also
indicate that the Dirichlet boundary condition reduce the
lightcone fluctuations that may already present in the Minkowski
spacetime, while the Neumann boundary conditions enhance them.


\subsection{Neumann boundary conditions}

After having considered the case of the Dirichlet boundary
conditions, here we turn our attention to the case in which
gravitons satisfy the Neumann boundary conditions. The
renormalized Hadamard function obeying the boundary conditions can
also be found, by using the method of images, as
\begin{eqnarray}
\label{eq:Neu}
G^R_{ijkl}(t,x,y,z;t',x',y',z')&=&\sum_{n=-\infty}^{+\infty}{'}G_{ijkl}^{(1)}(t,x,y,z;t',x',y',z'+2nL)\nonumber\\
&&+\sum_{n=-\infty}^{+\infty}G_{ijkl}^{(1)}(t,x,y,z;t',x',y',-z'+2nL)\;.
\end{eqnarray}
Note the sign change of the second term in the above expression as
compared to that in Eq.~(\ref{eq:2}).


\subsubsection{Parallel propagation case}


 First let us examine the
case in which a light ray propagates parallel to the planes from
point $(t,a,0,z)$ to point $(t',b,0,z)$. Using the same notation
and methods as before, we find that the mean squared geodesic
interval fluctuation $\langle\sigma_1^2\rangle_R$ can be expressed
as
\begin{eqnarray}
\label{eq:12}
\langle\sigma_1^2\rangle_R &=& \frac{r^2}{8}[f(2z,r)+f(2z-2L,r)\nonumber\\ &&+
\sum_{n=1}^{+\infty}(2f(2nL,r)+f(2z+2nL,r))+\sum_{n=2}^{+\infty}f(2z-2nL,r)]\;.
\end{eqnarray}
In the limit of $r\gg L$, the summation part in the above
expression can be calculated, in the same way as that in the
proceeding subsection, to get
\begin{eqnarray}
\label{eq:40}
&&\sum_{n=1}^{+\infty}(2f(2nL,r)+f(2z+2nL,r))+\sum_{n=2}^{+\infty}f(2z-2nL,r)\nonumber\\
&&\quad\quad\approx\frac{4}{3\pi^2}\left[3A^2\epsilon-5\ln
\epsilon +\left(1+\frac{z}{L}\right)\ln\left(1+{z\over L}\right)+
\left(2-\frac{z}{L}\right)\ln\left(2-{z\over L}\right)\right] \;,
\end{eqnarray}
where
\begin{eqnarray}
A^2= {1\over 4}\int_0^\infty dx \;{ \ln ( 2x+\sqrt{1+4x^2})\over
x\sqrt{1+4x^2}} \approx 0.6168\;.
\end{eqnarray}
Using Eq.~(\ref{eq:f}),  Eq.~(\ref{eq:f1}) and Eq.~(\ref{eq:40}),
we find that
\begin{eqnarray}
\langle\sigma_1^2\rangle_R\approx
\frac{r^2}{8}\left[\frac{4}{3\pi^2}\left(
3A^2\frac{r}{L}-3\ln\frac{r}{L}+\ln\frac{L^2}{z(L-z)}\right)\right]\approx\frac{A^2r^2}{2\pi^2}\frac{r}{L}\;.
\end{eqnarray}
Here in the last step we have assumed that $z\neq 0$ and $z\neq
L$. Thus
\begin{eqnarray}
\label{eq:tt} \Delta t\approx { A\over \sqrt{2}\pi}\sqrt{{r\over
L} }\;.
\end{eqnarray}
A comparison of Eq.~(\ref{eq:t}) and the above result shows that
while the growth of the mean deviation due to the lightcone
fluctuations over the travel distance, $r$,  is  that of a square
root of the logarithm of $r$ in the Dirichlet boundary condition
case, it is a square root of $r$ here. Therefore,  as $r$, the
travel distance increases,  the lightcone fluctuations grow much
faster in the present case than that in the corresponding case
with the Dirichlet boundary conditions. The square root behavior
of growth seems to suggest a fluctuation of random walk nature.
Note also that  this behavior is the same as that in the case
where one spatial dimension is periodically compactified, i.e.,
the space topology is that of a cylinder (see Eq.~(78) of
Ref.~\cite{yu1}), since the constant $A$ equals $c_1/2$ given by
Eq.~(75) in Ref.~\cite{yu1}.

 We now turn to the case in which  $L\gg r$ and $r\gg z$.
Here the summation part can be approximately calculated by series
expanding the summand first. The result is
\begin{eqnarray}
\sum_{n=1}^{+\infty}[2f(2nL,r)+f(2z-2nL,r)+f(2z+2nL,r)]\approx
\frac{2\pi^4}{45^2\times 21}\;{r^6\over L^6}\;.
\end{eqnarray}
Therefore, if in addition, $r\gg z$, we have
\begin{eqnarray}
\langle\sigma_1^2\rangle_R\approx\frac{r^2}{8}
\left[\frac{4}{3\pi^2}\ln\frac{r}{z}+\frac{2\pi^4}{45^2\times
21}\;{r^6\over L^6}\right]\;.
\end{eqnarray}
Thus
\begin{eqnarray}
\label{eq:ttt} \Delta t\approx
\sqrt{\frac{\ln(r/z)}{6\pi^2}+\frac{2\pi^4}{45^2\times
21}\;\left({r\over L}\right)^6}\;.
\end{eqnarray}
The leading term here is the same as that in the case of the
Dirichlet boundary conditions (refer to Eq.~(\ref{eq:tttt})). But
the higher order corrections are suppressed by a square factor of
$z/L$ in the Dirichlet boundary conditions as compared to that of
the Neumann ones. So again, the lightcone fluctuations here are
greater.  If we let $L\rightarrow \infty$, then the result of a
single plane is recovered as expected ( see Eq.~(91) in
Ref.~\cite{yu1}).
 However, if $r\ll z$, then we have
\begin{eqnarray}
\langle\sigma_1^2\rangle_R\approx\frac{r^2}{8}
\left[\frac{2\pi^4}{45^2\times 21}\;{r^6\over L^6}\right]\;.
\end{eqnarray}
As a result
\begin{eqnarray}
\label{eq:ttt1} \Delta t\approx{\pi^2\over
90\sqrt{21}}\;\left({r\over L}\right)^3 \;.
\end{eqnarray}
If we compare Eq.~(\ref{eq:t5}) with the above result, we can see
that the mean deviation in the present case  is larger than that
of the Dirichlet boundary conditions since $L\gg z$.


\subsubsection{Perpendicular propagation case}


Here we study the case in whch a light ray travels along $z$ axis
from point $(t,0,0,a)$ to $(t',0,0,b)$. Making use of
 Eq.~(\ref{eq:26}),  Eq.~(\ref{eq:27}),
 Eq.~(\ref{eq:30}), Eq.~(\ref{eq:31}), and Eq.~(\ref{eq:Neu}), we arrive at the
following results.

If $L\gg r$ and $r\gg a$, then
\begin{equation}
\langle\sigma_1^2\rangle_R\approx {r^2\over
8}\left[\frac{1}{3\pi^2}[1+\ln (r/a)]+{2r^2\over 9L^2}
+{\pi^2\over 45}\left({r\over L}\right)^2 \left({a\over
L}\right)^2\right]\;,
\end{equation}
and
\begin{eqnarray}
\Delta t=\frac{\sqrt{|\langle\sigma_1^2\rangle_R|}}
{r}\approx\sqrt{\frac{1}{24\pi^2}[1+\ln (r/a)]+{1\over
36}\left({r\over L}\right)^2}\;.
\end{eqnarray}
The $L$-independent terms in the above expression agree with the
result of a single plane (see Eq.~(85) in Ref.~\cite{yu1}), while
the $L$-dependent term gives the contribution due to the presence
of the second plane. The above result reveals that the lightcone
fluctuations are reinforced with the introduction of a second
plane. Recall Eq.~(\ref{eq:T1}), one can see that the
$L$-dependent term here is of lower order than that in the case of
the Dirichlet boundary conditions.

If, however, $r\ll a$, then
\begin{equation}
\langle\sigma_1^2\rangle_R\approx {r^2\over
8}\left[\frac{r^2}{9L^2} +{1\over 3}{r^2\over L^2}\csc^2(\pi
a/L)\right]\;,
\end{equation}
Thus
\begin{eqnarray}
\Delta t=\frac{\sqrt{|\langle\sigma_1^2\rangle_R|}}
{r}\approx\sqrt{\frac{r^2}{72L^2} +{1\over 24}{r^2\over
L^2}\csc^2(\pi a/L)}\;.
\end{eqnarray}
If we further assume that $ L\gg a$, then the above result becomes
\begin{eqnarray}
\Delta t\approx \sqrt{\frac{1}{24\pi^2}\left({r\over a}\right)^2
+{1 \over 36}\left({r\over L}\right)^2 }\;.
\end{eqnarray}
This shows that the lightcone fluctuations grow linearly with $r$
in this case, and the result reduces that of a single plane when
$L$ approaches infinity, i.e., one plane is infinitely away from
the other. Once again, the mean deviation is larger here than in
the case of the Dirichlet boundary conditions.


\section{Conclusion}


We have examined the quantum lightcone fluctuations in a spacetime
with two parallel planes based upon a framework proposed in
\cite{Ford1} and subsequently further developed in
\cite{Ford2,yu1,yu2}.  Lightcone fluctuations due to gravitons
satisfying the Dirichlet and the Neumann boundary conditions are
all considered.  For each boundary condition, the mean deviation
from the classical propagation time is calculated both for light
rays travelling parallel to the planes and for those propagating
perpendicularly to the planes. For all cases, we find that
lightcone fluctuations are larger when the Neumann boundary
conditions are imposed than that when the  Dirichlet boundary
conditions are enforced. This reveals that quantum lightcone
fluctuations depend not only on the geometry  and topology of the
spacetime as has been argued\cite{yu1} but also on the boundary
conditions.

 In the case of longitudinal (parallel) light propagation, when
the travel distance is large compared to the separation of the
planes, the fluctuation increases with the squared root of the
distance travelled for the case of the Neumann boundary
conditions, indicating a fluctuation of a random walk nature. It
is worth pointing out that this behavior is basically the same as
that in the corresponding case in a spacetime where one spatial
section is periodically compactified\cite{yu1}. However, if the
boundary condition is that of the Dirichlet, then the growth
becomes the square root of the logarithm  of the distance
travelled, which is much slower than the square root growth over
the travel distance in the Neumann boundary condition case.

The lightcone fluctuations discussed in this paper are
frequency-independent.  However, it is interesting to note that
frequency-dependent even helicity-dependent lightcone fluctuations
are expected in other theories of quantum gravity \cite{EMN99,GP}.
It should be pointed out that the growth of lightcone fluctuation
here is in general a rather complicated function of the light
travel distance and it can be linear, square root, etc., depending
on the relative size of the travel distance and the characteristic
scales of the configuration. Therefore contingent on the physical
scales involved, the effects of lightcone fluctuations obtained
here can either be larger or smaller than those effects discussed
in \cite{EMN99} using the same theoretic formalism together with a
string theory motivated microscopic model.

Finally, for all the cases we have investigated for both
longitudinal and transverse light propagations in both cases of
the Dirichlet and the Neumann boundary conditions, our results
show that lightcone fluctuations are greater in the spacetime with
two parallel planes than that in the spacetime with just a single
plane. Thus the presence of a second plane reinforces the quantum
lightcone fluctuations. This phenomenon  may be considered as  due
to enhanced quantum fluctuations stemming from confining gravitons
in a smaller region. It can be understood as a consequence of the
uncertainty principle and an analogy of what happens to any
quantum particle confined in a smaller region of space.

\begin{acknowledgments}
We would like to acknowledge the support  by the National Science
Foundation of China  under Grant No. 10075019, the support by the
Fund for Scholars Returning from Overseas by the Ministry of
Education of China and the support by the National Science
Foundation of Hunan Province  under Grant No. 02JJY2007.
\end{acknowledgments}


\end{document}